\documentclass[twocolumn]{jpsj2} %% two-column layout
\def\bm#1{{\mbox{\boldmath $#1$}}}

\def\bms#1{{\mbox{\rm\scriptsize\it\bf#1}}}

\def\a{\alpha}
\def\b{\beta}

\def\m{\mu}

\def\s{\sigma}

\def\G{\Gamma}

\def\bra{\langle}
\def\ket{\rangle}

\def\ua{\uparrow}
\def\da{\downarrow}

\def\nn{\nonumber}

\def\lsim{\mbox{\raisebox{-0.5ex}{$\stackrel{<}{\sim}$}}}

\title{Theory for Magnetic Anisotropy of Field-Induced Insulator-to-Metal Transition in Cubic Kondo Insulator YbB$_{12}$}

\author{Toshiyuki \textsc{Izumi}\thanks{Present address: Japan Radio Co. Ltd.}, Yoshiki \textsc{Imai} and Tetsuro \textsc{Saso}}

\inst{Division of Materials Sciences, Graduate Course of Science and Engineering, Saitama University, Shimo-Ohkubo 255, Sakura-ku, Saitama City, Saitama, 338-8570}

\abst{Magnetization and energy gap of Kondo insulator YbB$_{12}$ are calculated theoretically based on the previously proposed tight-binding model composed of Yb 5d$\epsilon$ and 4f $\Gamma_8$ orbitals.  It is found that magnetization curves are almost isotropic, naturally expected from the cubic symmetry, but that the gap-closing field has an anisotropy: the gap closes faster for the field in (100) direction than in (110) and (111) directions, in accord with the experiments.  This is qualitatively understood by considering the maximal eigenvalues of the total angular momentum operators projected on each direction of the magnetic field.  But the numerical calculation based on the band model yields better agreement with the experiment.}

\kword{Kondo insulator, YbB$_{12}$, magnetization, field-induced insulator-metal transition, magnetic anisotropy}

\begin{document}
\maketitle

\section{Introduction} 
YbB$_{12}$\cite{Kasaya85} is the most typical Kondo insulator, since it has a cubic crystal structure (Yb and B$_{12}$ forms NaCl structure), and many experiments: transport,\cite{Sugiyama88} calorimetric,\cite{Iga88} photoemission,\cite{Susaki96,Takeda06} optical,\cite{Okamura98,Okamura05} neutron scattering,\cite{Alekseev02,Bouvet98,Mignot05} etc. are already reported.  We believe that most of them can be understood in terms of the band insulator model with strong Coulomb repulsion between f electrons.  In fact, based on the LDA+U band calculation, it was found that the conduction bands near the Fermi level are composed of the 5d$\epsilon$ orbitals on Yb atoms, and the simple tight-binding band reproduces the LDA+U calculation rather well.\cite{Saso03}  Some of the conduction bands are doubly degenerate (four-fold if spin degeneracy is included).  A gap opens by the hybridization of the four-fold degenerate $\G_8$ states with the above-mentioned conduction bands with degeneracy.  It was pointed out that the gap can not open if the crystal-field groud state of 4f were $\G_7$ doublet.  Therefore, proper consideration of the degeneracy is a key to understanding the opening of a gap in the Kondo insulators.\cite{Saso03}

Based on this band model, one of the authors\cite{Saso04} calculated frequency- and temperature-dependence of the optical conductivity and obtained semi-quantitative agreement with the experiment.\cite{Okamura05}
He succeeded in reproducing the mid-infrared peak and the prominent temperature dependence of the low frequency part, which is due to the many-body effect.

In contrast to YbB$_{12}$, it is not clear whether other well-known or classical ``Kondo insulators'', SmS\cite{SmS} and TmSe,\cite{TmSe} belong to the same category as YbB$_{12}$, since Sm and Tm has more complicated 4f states than Yb.  In addition, TmSe has odd number of electrons per unit cell, so that it can not be a band insulator if 4f electrons are mixed valent.  Other Kondo insulators, which may belong to the same category as YbB$_{12}$, will be Ce$_3$Bi$_4$Pt$_3$\cite{Ce3Bi4Pt3} and Ce-skutterudites, e.g. CeFe$_4$As$_{12}$.\cite{CeFe4As12}  Nevertheless, YbB$_{12}$ can be a prototype of the Kondo insulator from the point of view of the simpleness of the band structure, sufficient number of experiments and the existence of the compact theoretical model.

Despite the above-mentioned success, YbB$_{12}$ still has many features to be understood correctly and quantitatively.  One of them is the magnetic properties.
Sugiyama, et al.\cite{Sugiyama88} first found that the activation energy of the resistivity of this insulator vanishes at aroud 50 Tesla when magnetic field is applied.  This is called as field-induced insulator-to-metal transition and the critical field is denoted as $B_c$ hereafter.  Iga, et al.\cite{Iga99} found that the magnetization curve for each direction does not show an anisotropy, but the critical field $B_c$ does.  Namely, $B_c(110)$ and $B_c(111)$ is larger than $B_c(100)$ by 10 percent ($B_c(100)=$47T, $B_c(110) \simeq B_c(111)=$53T).  The transitions are first order since hysterisis is observed.

Using the simple periodic Anderson model, one of the present authors showed that the overall feature of the magnetization curve of YbB$_{12}$ can be explained and the transition can be first order if the Coulomb repulsion is taken into account.\cite{Saso96,Saso97}  In addition, he found that the renormalization factor (which renormalizes the gap) due to strong correlation does not largely change by the magnetic field in the insulating phase, so that the renormalized up- and down-spin bands are shifted by the field as if they are rigid bands and the renormalized band gap is closed also as if in the case of the rigid bands.\cite{Saso96}  But the anisotropic feature of the f-electrons under cubic crystalline field was not taken into account.

In the present study, we investigate theoretically the origin and mechanism of the anisotropy of the field-induced insulator-to-metal transition.  In \S 2, we briefly explain the previously introduced tight-binding band model.  Calculations of magnetization and energy gap under magnetic field will be presented in \S 3.  The last section is devoted to the discussions and conclusions.

\section{Tight-Binding Energy Band}

The energy dispersions of the conduction bands of YbB$_{12}$ are well approximated by the tight-binding model of 5d$\epsilon$ orbitals of Yb on the fcc lattice and expressed by the following simple expressions:\cite{Saso03}
\begin{equation}
  E_\bms{k}^{\alpha\beta} = E_{{\rm d}\varepsilon}+3({\rm dd}\sigma)\cos(k_\alpha/2)\cos(k_\beta/2),
  \label{eq:dxy-band}
\end{equation}
where (dd$\s$) is a Slater-Koster integral,\cite{Slater54} $(\a,\b)$ denotes $(x,y)$, $(y,z)$ or $(z,x)$.
The lowest band along $\Gamma$-K-X(110) and $\Gamma$-X(100) are doubly degenerate (four-fold if the spin degeneracy is included).
Introducing the hybridization between these d bands and the 4f $j=7/2$ $\Gamma_8$ states, we obtained the energy bands with a gap.

In eq.(\ref{eq:dxy-band}), we locate the energy level of 5d$\varepsilon$ orbitals at $E_{\rm{d}\varepsilon}$=1.0 Ry, and set (dd$\sigma$)= 0.06 Ry.
This value of (dd$\sigma$) is the effective one due to hopping between near-neighbor Yb sites through B$_{12}$ clusters.

The hybridization between these 5d bands and the 4f states is described by the effective (df$\sigma$) integrals\cite{Takegahara80} between the nearest-neighbor Yb sites.
We retain only the nearest neighbor (df$\sigma$) bonds as the simplest model.  
We locate the 4f $\Gamma_8$ states at $E_{\Gamma_8}=0.88$ Ry and choose (df$\sigma$)=0.015 Ry.
We also include (df$\pi)=-0.0075$ Ry and (ff$\sigma)=-0.003$ Ry to adjust to the LDA+U calculation.  Furthermore, the filled bands below the gap are shifted down by $\Delta E=-0.011$ Ry relative to the bands above the gap.  This treatment is in accord with the spirit of the LDA+U treatment.
These parameters are the same as those in our previous paper\cite{Saso04} to fit the optical conductivity.\cite{Okamura05}  

Using these parameters, we obtain the dispersion curves which reproduce the LDA+U bands around the gap rather well.\cite{Saso03} They have an indirect gap of about 0.0069 Ry ($=$ 1089 K) between X and L points and the direct gap of 0.018 Ry ($=$ 0.245 eV). 
The former is about six times larger than the best estimated value of the indirect gap in the optical experiment, $E_g=$174 K.\cite{Okamura05}  However, in the present realistic band model, we can not use this value. If we use a band model with the correct value of the indirect gap, it is difficult to reproduce the gap in the density of states, since we must use a very fine mesh to calculate the three-dimensional $\bm{k}$-summation over the Brillouin zone.

Note that the band calculation can treat the f-state only by the $j$-$j$ coupling scheme.
One hole in $j=7/2$ $\Gamma_8$ state corresponds to the $\Gamma_8$ state with the total angular momentum $J=7/2$ of Yb$^{3+}$ (the f-electron number $n_f=13$).
In the present case, the number of filled 4f $\Gamma_8$ electrons is about 3.1 and 0.9 hole exists in the bands above the gap, whereas $j=5/2$, $j=7/2$ $\Gamma_6$ and $\Gamma_7$ states are far below $\Gamma_8$ and are implicitly treated as completely filled.  Thus the present model approximately reproduces the experimental valence of Yb$^{+2.9}$.

\begin{figure}[t]
\begin{center}
\includegraphics[width=8cm]{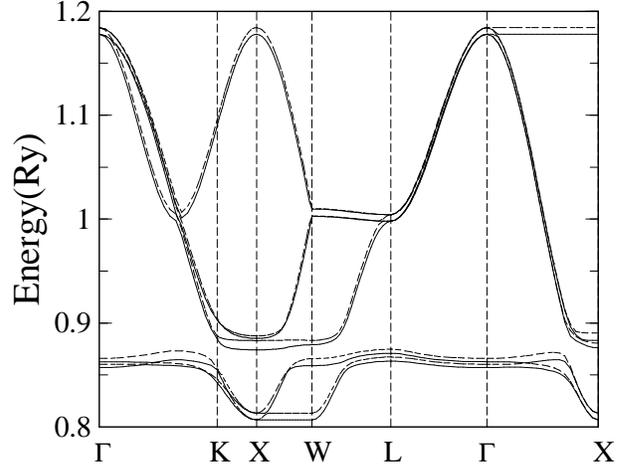}
\end{center}
\caption{The tight-binding band for YbB$_{12}$ under B=500T in (100) direction.  The X points in the left and right denote the equivalent (110) and (100) points (in unit of $2\pi/a$), respectively.}
\label{Fig:Band}
\end{figure}

\section{Calculation of Magnetization and Energy Gap under Magnetic Field}
We apply the magnetic field $\bm{B}=\m_0\bm{H}$ along (100), (110) and (111) directions.  The Hamiltonian reads

\begin{equation} 
{\cal H}=
  \sum_{\stackrel{\alpha,\beta}{i,j}} t_{ij}^{\alpha\beta} c_{i{\alpha}}^{\dagger}c_{j{\beta}}
 +\sum_{i}g_{c}\mu_{B}\bm{s}_i\cdot\bm{H}
 +\sum_{i}g_{J}\mu_{B}\bm{j}_i\cdot\bm{H},
\end{equation} 
where $t_{ij}^{\alpha\beta}$ denotes the hopping matrix elements between the ten orbitals $\{ \a,\b= xy\ua, yz\ua, zx\ua, xy\da, yz\da, zx\da, \G_8^{(1)\pm}, \G_8^{(2)\pm} \}$ on $i$ and $j$ sites and is expressed by the Slater-Koster parameter given in the last section.\cite{Saso04}  $\m_B$ is the Bohr magneton, $g_c=2$ and $g_f=8/7$ are the g-factors of conduction and f electrons.  $\bm{s}_{i}$ and $\bm{j}_{i}$ are the spin and the total angular momentum operators for the conduction and f electron at $i$ site, respectively.  We have omitted the effect of magnetic field on the orbital motion except that through $\bm{j}$.  The effect of orbital motion can be included by attaching the phase factor to the hopping matrix elements, but it will be smaller than those considered here through $\bm{j}$, since the latter will be enhanced by the strong correlation.  

In the present model with the larger gap than the experiment, the gap closes at the large field of about 500 T.  We show the band structure at this field in (100) direction in Fig.\ref{Fig:Band}.  Four bands lie below the gap and six bands above the gap and all the degeneracies under zero field are lifted.

Diagonalizing the Hamiltonian, the Bloch function is obtained as
\begin{equation}
  \psi_{\bms{k}n}(\bm{r}) = \frac{1}{\sqrt{N}} \sum_{i,\a} u_{\bms{k}n,\a}e^{i\bms{k}\cdot\bms{R}_i}\phi_\a(\bm{r}-\bm{R}_i),
\end{equation}
where $\phi_\a(\bm{r}-\bm{R}_i)$ is the local orbital $\a$ at the site $\bm{R}_i$, $n$ denotes the diagonalized band index, $u_{\bms{k}n,\a}$ the coefficient and $N$ the total number of unit cell.
Using this, the magnetization $\bm{M}$ is calculated by
\begin{eqnarray}
  \bm{M} & = & \frac{1}{N} \sum_{\bms{k},n} f(E_{\bms{k}n}) \left[ g_c\m_B\bra\psi_{\bms{k}n}|\bm{S}|\psi_{\bms{k}n}\ket \right. \nn \\
        & & + \left. g_f\m_B\bra\psi_{\bms{k}n}|\bm{J}|\psi_{\bms{k}n}\ket \right],
\end{eqnarray}
where $\bm{S}=\sum_i\bm{s}_i$, $\bm{J}=\sum_i\bm{j}_i$, $f(E)$ is the Fermi function and $E_{\bms{k}n}$ the $n$-th band energy.
Note that the application of the magnetic field lowers the symmetry, so that the sum over 1/48 of the first Brillouine zone is not allowed.  
In the actual calculation, we summed over whole of the first and the second Brillouine zones because the programming becomes much easier: we can simply make summation over a cube with twice the volume of the first Brillouine zone of fcc lattice.  We divide this cube into at most $400^3$ mesh points.  Temperature is set equal to zero throughout the paper.  

Also for simplicity,  we have summed only over the lowest four filled bands in the calculation of the magnetization, so that the calculation makes sense only for $B<B_c$ (critical field) in each field direction.  For $B>B_c$, one has to determine the chemical potential for each value and direction of the field and sum up over more than four bands.  We did not do it since we are interested only in the anisotropy problem in the present paper.

\begin{figure}[htbp]
  \begin{center}
    \includegraphics*[width=8cm]{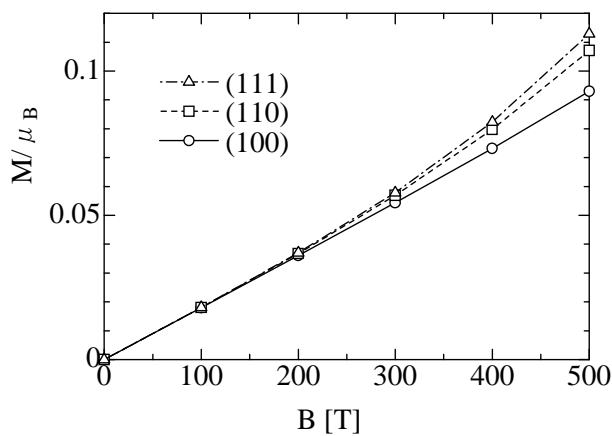}
    \caption{Magnetization curves under the magnetic field in (100), (110) and (111) directions.}
    \label{Fig:Mag}
  \end{center}
\end{figure}

\begin{figure}[htbp]
  \begin{center}
  \vspace{2cm}
    \includegraphics*[width=8cm]{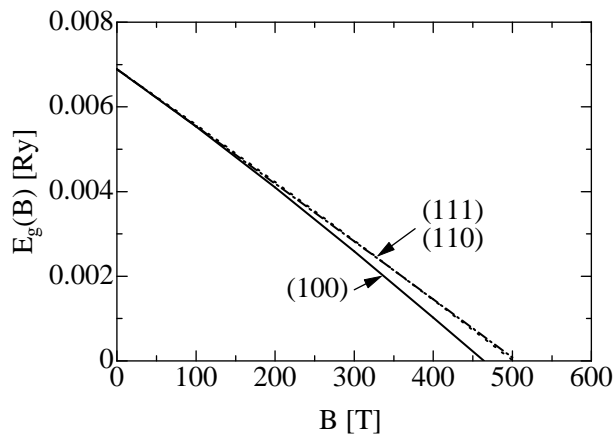}
    \caption{Closing of the energy gap by the magnetic field in (100), (110) and (111) directions.  Difference between (110) and (111) curves is almost invisible.}
    \label{Fig:Gap}
  \end{center}
\end{figure}

$M/\mu_B=|\bm{M}|/\mu_B$ is plotted in Fig.\ref{Fig:Mag} for the three directions of the field.
In accord with the general theorem of the group theory that the linear response tensor must be proportional to the unit tensor (isotropic), the present result shows that the magnetization is isotropic up to 200 T.  
(Indirect) energy gap vs. magnetic field is plotted in Fig.\ref{Fig:Gap}.  Since we take large value of the initial gap, the critical value of the field in each direction is also large: $B_c(100)=464$T, $B_c(110)=501$T, $B_c(111)=506$T.  These values will become smaller if we use the correct value of the band gap.  For $B<300$T, $E_g(111)$ becomes slightly smaller than $E_g(110)$, but the difference is very small and it seems out of range of the precision of the present calculation.  It seems reasonable to consider that $E_g(100) < E_g(110) \lsim E_g(111)$ for all the field, as will be discussed in the next section.

\section{Discussions and Conclusions}
\begin{figure}[htbp]
  \begin{center}
    \includegraphics*[width=7cm]{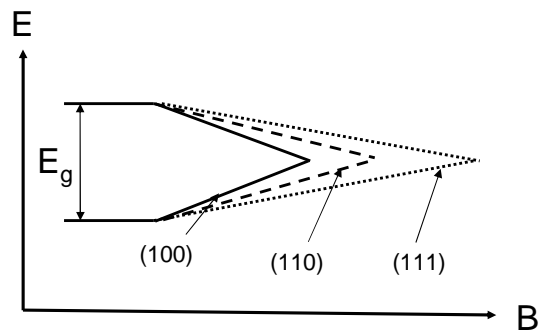}
    \caption{Schematic figure to explain how the gap is closed when the magnetic field is applied in each direction.}
    \label{Fig:GapClosing}
  \end{center}
\end{figure}

In order to understand an origin of the anisotropy of the critical field, we define the angular momentum operators projected onto each direction: $\bm{j}_{100}=\bm{j}\cdot\bm{u}_{100}=j_x$, $\bm{j}_{110}=\bm{j}\cdot\bm{u}_{110}=(j_x+j_y)/\sqrt{2}$, $\bm{j}_{111}=\bm{j}\cdot\bm{u}_{111}=(j_x+j_y+j_z)/\sqrt{3}$, where $\bm{u}_{100}$, $\bm{u}_{110}$ and $\bm{u}_{111}$ are the unit vectors in each direction.  Making the matrix representation of these operators in the $\Gamma_8$ space and diagonalize the 4 $\times$ 4 matrices, we obtain the maximal eigenvalue for each direction as follows: 1.833 in (100), 1.635 in (110) and 1.500 in (111).  From this result, one can understand that the f-states above and below the gap is shifted by the field faster in this order. (See Fig.\ref{Fig:GapClosing}.)  There is, however, a difference between the above explanation and the experiment.  Namely, the critical fields in (110) and (111) directions are almost the same, whereas $B_c(100)$ is 10 per cent smaller in the experiment.  Our numerical calculation based on the realistic band model presented in Fig.\ref{Fig:Gap}. successfully reproduces this experimental feature.  It may be because the percentage of each $\Gamma_8$ component is different at each $\bm{k}$ point, so that the above-mentioned explanation using the maximal eigenvalues of projected angular momentum operators holds only qualitatively.

\vspace{2cm}

The other problem is that the anisotropy of the gap-closing is not clearly visible in Fig.\ref{Fig:Gap} in the low field region where the magnetization is isotropic ($B<200$ T).  This may be partly because of insufficiency of the number of the $\bm{k}$-mesh points.  We anticipate, however, that the clear anisotropy observed in the high field range in Fig.\ref{Fig:Gap} is a real one and may survive also in the low field region in accord with the experiment.

One more problem in our theory is that the calculated magnetization is too small.  Value of the magnetization right before the metal-to-insulator transition is about 0.3 $\sim$ 0.4 $\mu_B$/Yb in the experiment,\cite{Iga99} but the calculation shows only about 0.1 $\mu_B$/Yb even at 500 T, and $M/\mu_B\sim 0.01$ at $H=50$ T.  Namely, about 30$\sim$40 times enhancement is necessary.  It is clear that the slope of the magnetization curve will be enhanced by a many-body effect,\cite{Saso96} but a quantitative calculation using the present model with full anisotropy is not easy.  Nevertheless, it was already clarified that the renormarization factor does not strongly depend on the magnetic field until the gap is closed by the field at least for the simple periodic Anderson model\cite{Saso96} as mentioned in the Introduction.   Since the magnetization must be isotropic in the linear region, the enhancement factor must also be isotropic.  Thus, the many-body effect will give only a quantitative effect on the slopes of the magnetization curves and may not affect the anisotropy of the critical field.

In conclusion, using the tight-binding model composed of Yb 5d$\epsilon$ and 4f $\Gamma_8$ orbitals, which reproduces the LDA+U band structure near the energy gap, we have calculated the magnetization curves and the collapse of the energy gap of the Kondo insulator YbB$_{12}$ when the magnetic field is applied in (100), (110) and (111) directions.  Since the initial gap was taken several times larger than the experimental one due to a technical reason in the numerical calculation, the critical values of the fields at which the gap vanishes are too large. We, however, have successfully explained the order of the direction of the critical field: $B_c(100)<B_c(100)\approx B_c(111)$.  Although many-body effect is not included, we have disscussed that the results may not change even if the strong correlation is taken into account.  Of course, more perfect calculation should be performed including correlation and a small initial gap, but we believe that the present study can be an important step to understand the properties of the Kondo insulators from the microscopic point of view, especially, from the model based on the LDA(+U) band calculation plus correlation.

\section*{Acknowledgment}
One of the authors (T. S.) acknowledges F. Iga for correspondense on experiments.

\end{document}